\documentclass[conference]{IEEEtran}
\IEEEoverridecommandlockouts

\usepackage{amsmath, amsfonts}
\usepackage{algorithmic}
\usepackage{graphicx}
\usepackage{textcomp}
\usepackage{xcolor}

\usepackage{fancyhdr}
\usepackage{url}

\usepackage{subfig}
\usepackage{textcomp}
\usepackage{tabularx}%
\usepackage{booktabs} %
\usepackage{fancyhdr}
\usepackage{microtype}      %
\usepackage{multirow}
\usepackage[labelfont=bf, skip=0pt]{caption}
\usepackage{xcolor, colortbl}
\usepackage{multicol}
\usepackage{physics}
\usepackage{makecell}
\usepackage{float}
\usepackage{mathtools, nccmath}
\usepackage[ruled,vlined,linesnumbered,noresetcount]{algorithm2e}
\usepackage{diagbox}
\usepackage{footmisc}
\usepackage{adjustbox}
\usepackage{mwe}
\usepackage{pifont}%
\usepackage{cite}
\usepackage{hyperref}

\usepackage{tikz}

\newcommand{\name}{GuardNN\xspace}
\newcommand{\versionnumber}{version number}
\newcommand{\VN}{VN\xspace}
\newcommand{\VNs}{VNs\xspace}

\definecolor{ForestGreen}{rgb}{0.13, 0.55, 0.13}

\usepackage{tabu}
\usepackage{cases}
\usepackage{comment}

\definecolor{brickred}{rgb}{0.8, 0.25, 0.33}
\definecolor{forestgreen}{rgb}{0.13, 0.55, 0.13}

\begin{document}

\title{GuardNN: Secure Accelerator Architecture for Privacy-Preserving Deep Learning}

\author{
\IEEEauthorblockN{Weizhe Hua$^{\dagger}$, Muhammad Umar$^{\dagger}$, Zhiru Zhang$^{\dagger}$, G. Edward Suh$^{\dagger\mathsection*}$}
\thanks{$^*$Work was done while at Cornell University.}
\IEEEauthorblockA{\textit{$^{\dagger}$Cornell University, Ithaca, NY, USA}\\{\textit{$^{\mathsection}$Meta AI, Cambridge, MA, USA}}}\\
\textit{\{wh399,mu94,zhiruz,gs272\}@cornell.edu, edsuh@fb.com}
}

\maketitle

\begin{abstract}
This paper proposes GuardNN, a secure DNN accelerator that provides hardware-based protection for user data and model parameters even in an untrusted environment. GuardNN shows that the architecture and protection can be customized for a specific application to provide strong confidentiality and integrity guarantees with negligible overhead. The design of the GuardNN instruction set reduces the TCB to just the accelerator and allows confidentiality protection even when the instructions from a host cannot be trusted. 
GuardNN minimizes the overhead of memory encryption and integrity verification by
customizing the off-chip memory protection for 
the known memory access patterns of a DNN accelerator. 
GuardNN is prototyped on an FPGA, demonstrating effective confidentiality protection with $\sim$3\% performance overhead for inference.
\end{abstract}

\vspace{-0.05in}
\section{Introduction}
\label{sec:intro}
The past decade has seen unprecedented growth in the use of machine learning (ML). 
However, the data-intensive nature of ML raises serious concerns for security and privacy. 
Deep neural networks (DNNs) require collecting, storing, and processing a large amount of personal and private user data.
Moreover, 
DNN computations are often performed in mobile or cloud environments where private data may be exposed or misused.
For large-scale deployment of DNNs in privacy-sensitive applications, we need a way to perform DNN computations even in an untrusted environment, with both high performance and strong privacy protection.

A promising approach for providing strong confidentiality and integrity guarantees under untrusted environments
is to create a hardware-protected trusted execution environment (TEE), also called an enclave as in Intel SGX \cite{sgx}. 
So far TEEs have primarily been studied in the context of general-purpose processors, which 
cannot provide enough performance and energy efficiency for large-scale ML workloads.
This paper proposes to leverage application-specific accelerators to enable high-performance TEEs for ML, and
presents a secure DNN accelerator architecture, named \textit{\name}.

To protect sensitive data, \name~keeps
all confidential information including inputs, outputs, training data, and network parameters (weights) in an encrypted form outside of the trusted hardware boundary, such as an ASIC accelerator chip or an accelerator IP in an SoC.
Each accelerator contains a unique private key that can only be used by the accelerator itself. 
Users can remotely authenticate the accelerator using the corresponding public key and the certificate, and send private inputs and weights encrypted to the accelerator.
In this way, \name can ensure that an adversary cannot access private user data and weights even if he/she controls software or has physical access to the accelerator.
The secure accelerator can also protect the integrity of ML computation by incorporating remote attestation and off-chip integrity verification.

While {\name} adopts the high-level approach of today's TEEs, an accelerator TEE needs to address the challenge of providing protection while allowing both a CPU and an accelerator to work together.
The accelerator TEE also presents an opportunity to customize protection for a specific application domain to 
improve both security and performance. 
For example, while processors can perform arbitrary operations and memory accesses, accelerators only need to support a relatively small set of operations
and often have a memory access pattern that is specific to the target application. 
This application-specific nature of accelerators enables {\name} to customize its architecture
and protection mechanisms for ML, and provide strong security with almost no 
performance overhead.

The following summarizes the key benefits and insights that {\name} provides 
compared to today's general-purpose TEE: %
1) \name~carefully designs its architecture and instructions to enable confidentiality and integrity
protection even when a host CPU that controls scheduling and resource allocation cannot be trusted. This design
reduces the trusted computing base (TCB) to just the accelerator.
2) The \name~instruction set allows confidentiality-only protection, which is sufficient for
privacy-preserving ML, without the complexity and overhead of integrity protection.
Regardless of the sequence of instructions, private data are always encrypted outside the accelerator.
3) The accelerator TEE has the potential to provide stronger security compared to CPU TEEs;
an accelerator is physically separated from general-purpose cores and has much simpler hardware and software. 
The memory access pattern and the timing of a DNN accelerator without dynamic pruning is also independent of input and weight values, making \name{} secure against memory and timing side channels.

We implemented a prototype system based on CHaiDNN \cite{chaidnn}, an open-source DNN accelerator from AMD Xilinx. 
The experiments on an AMD Xilinx FPGA demonstrate functional correctness and show that the overhead of memory encryption is negligible.
For more detailed analyses, we performed experiments using cycle-level simulations.
The simulation results show that \name~can guarantee both confidentiality and integrity with small overhead. 

\vspace{-0.05in}
\section{Secure Accelerator Architecture}
\label{sec:arch}

\subsection{Threat Model}
\label{sec:threat_model}

We assume that a DNN accelerator can run both inference and training.
A scheduler runs on a host CPU and coordinates compute and data movement by communicating with a remote user and issuing commands to the DNN accelerator.
The remote user sends inputs and a DNN model, and receives final results.

The goal of a secure DNN accelerator is to protect the confidentiality and the integrity of DNN data and computation in an environment where only the accelerator itself can be trusted.
For confidentiality, the secure DNN accelerator aims to protect inputs (inference inputs or training data), prediction results, network parameters, and all intermediate results.
On the other hand, we consider the DNN structure as public information and do not hide the structure.
For integrity, the secure DNN accelerator aims to detect any unauthorized changes to its state and execution so that a user can verify that the output is the outcome of the given model/input.

The DNN accelerator is trusted and authenticated by the remote user using a unique private key that is only known by the accelerator hardware.
The accelerator needs to be designed and fabricated by a trusted manufacturer. The manufacturer also needs to securely embed a private key specific to each accelerator instance, and provide a certificate.
We assume that the internal operations and state of the DNN accelerator cannot be directly observed or changed 
by an adversary whereas anything outside of the accelerator including off-chip memory and a host processor 
are assumed untrusted.

A typical DNN model has a fixed memory access pattern, and the timing for a given model is agnostic to inputs and weights. %
In that sense, the \name~ accelerator is secure against memory and timing side-channel attacks. 
We do not consider other physical side-channel attacks such as the power and EM side channels.

\vspace{-0.05in}
\subsection{Key Insights and Features}
{\bf Small TCB:} DNN accelerators rely on a complex ML software stack for optimizations, scheduling, and resource allocation decisions. 
Most accelerators cannot efficiently run complex code and need to rely on a CPU for commands.
A straightforward design is to protect both CPU and a DNN accelerator in a TCB, where trusted software runs inside a TEE on the host CPU and controls TEE on the accelerator.
This design requires not only TEEs on both the CPU and the accelerator, but also a secure
communication channel between the two (possibly from multiple vendors), and a remote attestation
mechanism that allows users to verify the trustworthiness of a combination of multiple hardware components.
In that sense, directly extending today’s CPU TEE will lead to a large TCB with complex mechanisms.

Although the ML software stack is complex, the instruction set architecture (ISA) of a DNN accelerator remains simple because the DNN operations can be boiled down to scalar, vector, matrix additions and multiplications and a limited number of non-linear functions.
For example, TPU-v1~\cite{google2017tpu} only has a dozen instructions with five important ones.
Based on this observation, we propose to run the ML software on an untrusted host, while restricting the host interface to a limited set that does not leak sensitive information.

\name~can ensure confidentiality without trusting a host processor by designing its ISA so that sensitive information is always encrypted no matter which instruction is executed.
The outputs are encrypted and can be decrypted only by the remote user who owns the input and the model.
Leveraging the typical nature of DNN computations, {\name} also ensures that the latency of each instruction 
is independent of secret values.
The untrusted host chooses which DNN operations to be performed, but cannot make the accelerator produce outputs in plaintext. 
The \name~design significantly reduces the size of the TCB while allowing the flexibility and performance optimizations provided by the ML software.

{\bf Confidentiality-Only Protection:} 
{\name} can decouple confidentiality and integrity
protection, and protect the confidentiality of private data without paying the cost of integrity protection.
Regardless of the sequence of instructions, private data are always encrypted outside the accelerator.
In addition, the memory access patterns and execution times of DNN accelerators without dynamic pruning~\cite{cnvlutin, cgnet, cgnet_nips} are independent of input data values.
Hence, the confidentiality guarantees of \name~do not depend on the integrity of the instruction sequences and data values.
In contrast, the CPU TEEs require integrity protection even for confidentiality;
because the trusted software inside the TEE is allowed to output confidential information unencrypted, 
the integrity of the program must be protected even when only confidentiality is needed.

{\bf DNN-Specific Memory Protection:} 
Leveraging the regular and coarse-grained data movement patterns of DNNs, 
{\name} removes the need to store version numbers for memory encryption and integrity verification in off-chip memory. 
This DNN-specific optimizations enable protection with negligible overhead.

\begin{figure}[!t]
  \begin{center}
        \includegraphics[scale=0.41]{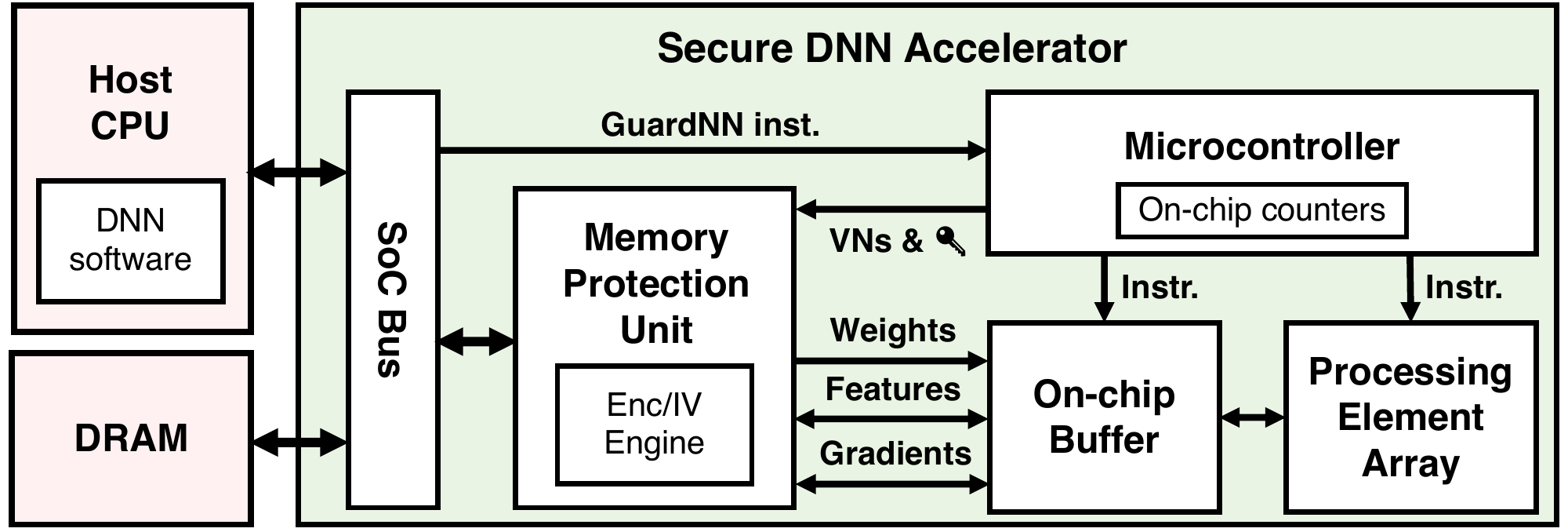}
        \vspace{0.1cm}
        \caption{\name~architecture overview --- \small{\normalfont{The green and red boxes represent trusted and untrusted components, respectively.}}} %
        \vspace{-0.3cm}
   \label{fig:arch}
  \end{center}
\end{figure}

\vspace{-0.05in}
\subsection{\name~Architecture}

\begin{table}[]
\centering
\caption{The security features provided by \name.}
\label{tab:security_analysis}
\begin{adjustbox}{width=\linewidth}
\large

\begin{tabular}{@{}lll@{}}
\toprule
\addtolength{\tabcolsep}{-2pt}
\large
\begin{tabular}[c]{@{}l@{}}Security Function\end{tabular} & Mechanism & Threat \\ \midrule
\textbf{Key Generation} & True random number generator & Replay/key guessing\\\midrule
\textbf{Key Exchange} & DHE key-exchange protocol & Untrusted host/network\\\midrule
\begin{tabular}[c]{@{}l@{}}\textbf{Off-chip Mem.}\\\textbf{Protection}\end{tabular} & \begin{tabular}[c]{@{}l@{}}DNN-optimized memory \\encryption and integrity verification \end{tabular} & \begin{tabular}[c]{@{}l@{}} Untrusted host/\\physical attacks\end{tabular}\\\midrule
\begin{tabular}[c]{@{}l@{}}\textbf{Restricted}\\\textbf{Instruction Set}\end{tabular} & \begin{tabular}[c]{@{}l@{}}No instruction allows outputting\\secrets in plaintext \end{tabular} & Untrusted host\\\midrule
\begin{tabular}[c]{@{}l@{}}\textbf{Remote}\\\textbf{Attestation}\end{tabular} & \begin{tabular}[c]{@{}l@{}}Hashes of input, output, weights, \\and instructions; Sign the hashes\end{tabular} & Untrusted host\\\midrule
\begin{tabular}[c]{@{}l@{}}\textbf{Side-channel}\\\textbf{Protection}\end{tabular} & \begin{tabular}[c]{@{}l@{}}The memory access pattern and the\\ timing are independent of secrets\end{tabular} & \begin{tabular}[c]{@{}l@{}}Memory and timing\\side-channels\end{tabular}\\
 \bottomrule
\end{tabular}
\end{adjustbox}
\vspace{-0.05in}
\end{table}

Here we introduce the {\name} architecture and its protection mechanisms.
Figure~\ref{fig:arch} shows the high-level block diagram, and 
Table~\ref{tab:security_analysis} summarizes the protection mechanisms.

The accelerator needs to be able to establish a secure communication channel with a remote user. %
For this purpose, a {\name} accelerator includes a unique private key (SK$_\text{Accel}$), a true random number generator, and and a microcontroller.
We assume that the user obtains the corresponding public key using a public key infrastructure as in Intel SGX or Trusted Platform Modules (TPMs).
{\name} also introduces an instruction that allows the accelerator to securely exchange a symmetric key (K$_\text{Session}$) and establish a secure communication channel
with the remote user.
The user sends DNN model weights, inputs, and outputs through the secure communication channel.
{\name} provides instructions to import encrypted inputs and weights, and produce an encrypted output.

During the execution, {\name} protects data in external memory using
a memory encryption (Enc) engine that 
encrypts data in DRAM, and an integrity verification (IV) engine that detects
unauthorized changes on a read from DRAM.
To minimize the performance overhead of memory protection, {\name} uses a DNN-specific memory protection.

The computation on a DNN accelerator is typically controlled by a software
scheduler on a host CPU.
While both CPU and accelerators can be protected by a TEE as in a recent GPU TEE design~\cite{graviton},
{\name} enables protection even when the host CPU cannot be trusted.
The instruction set is carefully designed to ensure that confidential information
is always encrypted outside the accelerator no matter which instruction runs.
For side-channel protection, {\name} requires that the timing and memory access pattern of the accelerator are independent of secret data.
This design ensures that confidentiality is protected independent of an instruction sequence. %
For integrity, {\name} computes the hashes of inputs and weights when they are imported, and keeps the hash of the sequence of executed instructions and their input arguments, similar to how remote attestation maintains the hash for software state.
Then, {\name} provides an instruction that signs the hashes of each output with the DNN data and instructions using the accelerator's private key so that a user can verify the initial state and the execution.

\vspace{-0.05in}
\subsection{Off-chip Memory Protection}
\label{sec:low_overhead_mp}

\subsubsection{Memory Protection Basics}

The counter-mode encryption (AES-CTR) is widely used in secure processors~\cite{cpu_MEE} to hide AES latency.
AES-CTR requires a non-repeating counter value for each encryption, which consists of the physical memory address (PA) of the data block that will be encrypted and a (per-block) {\versionnumber} (VN) that is incremented on each memory write.
To prevent data being tampered with by an attacker, integrity verification calculates and stores the MAC of the data value, PA and VN for each data block and checks that MAC on the following read. %
In addition, to defeat the replay attack, a Merkle tree (i.e., hash tree) \cite{hash_tree} is used to
verify the MACs hierarchically in a way that the root of the tree is stored on-chip.

\subsubsection{DNN-specific Protection}
\begin{figure}%
   \subfloat[Inference.]{
      \begin{minipage}{\linewidth}
      \centering
      \includegraphics[scale=0.3]{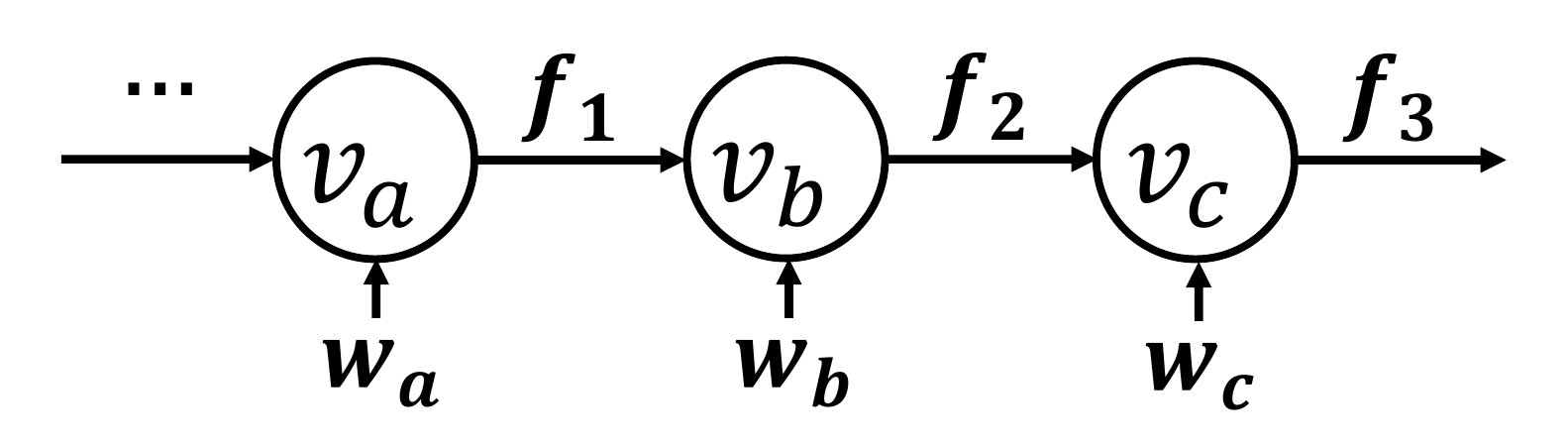}\\
     \label{fig:cnn-inf}
   \end{minipage}}
   \hspace{0.1cm}
   \subfloat[Training.]{
   \begin{minipage}{\linewidth}
   \centering
     \includegraphics[scale=0.3]{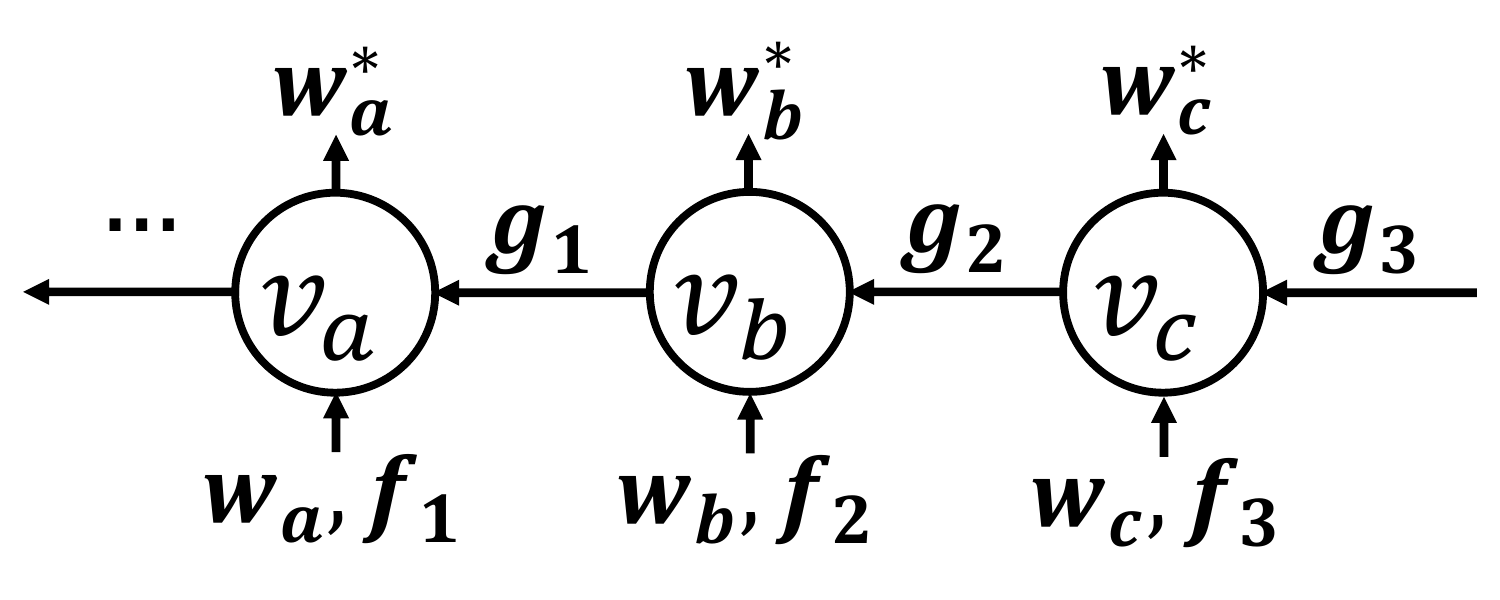}\\
     \label{fig:cnn-train}
   \end{minipage}}
\caption{DNN data-flow subgraphs.}
\vspace{-0.1cm}
\label{fig:dataflow}
\end{figure}

The main overhead of memory protection comes from accessing the off-chip \VNs~and MACs.
Because DNN accelerators are often memory-intensive, the additional meta-data accesses can lead to a non-trivial slowdown.
Popular ML frameworks often represent the network as a static data-flow graph (DFG) as illustrated in Figure~\ref{fig:dataflow} and optimize the graph before execution. 
Unlike CPUs, DNN accelerators have the same access pattern to a large chunk of memory.
A DNN accelerator typically reads/writes the output features of a layer (e.g., $f_1, f_2, f_3$ in Figure~\ref{fig:cnn-inf}) from/to DRAM the same number of times.
This regular memory access pattern allows using only one VN for all output features of a layer.
For example, if DNN accelerator only writes the features to DRAM once per layer, the layer number can be used as part of the VN.

In \name, each counter value for memory encryption includes the address of the 128-bit memory block being encrypted/decrypted and a 64-bit VN, and is used as the input to AES-CTR encryption. The VNs are constructed using a few on-chip counters to ensure that the counter values are unique for each encryption.
For writing new features, we introduce CTR$_\text{IN}$ and CTR$_\text{F,W}$ in the accelerator state to keep track of the number of inputs and the number of times that features are written for the current input.
CTR$_\text{IN}$ is incremented for each new input ({{\tt \textbf{SetInput}}}).
CTR$_\text{F,W}$ is reset on a new input and incremented after each DNN computation instruction ({\tt \textbf{Forward}}) that writes output features. 
The VN for writing features includes both CTR$_\text{IN}$ and CTR$_\text{F,W}$.

As the host CPU owns the DFG and controls the scheduling of instructions, the host CPU can easily reconstruct the VN used to write features.
For reading the features, \name~uses CTR$_{\text{F,R}}$ from the CPU to form the VN, and thus avoids tracking the status of the DNN tasks.
CTR$_{\text{F,R}}$ corresponds to the value of CTR$_\text{F,W}$ used to write the features.
As CTR$_{\text{F,R}}$ is only used in decryption, the confidentiality is not broken even if the CTR$_{\text{F,R}}$ value is incorrect.

The weights are read-only during inference. Therefore, we can use a constant as the VN for the weights.
To allow updating weights during training, {\name} keeps CTR$_{\text{W}}$ in the accelerator state and
keeps track of the number of updates to the weights ({\tt \textbf{SetWeight}}).
During training, each gradient edge in the DFG has a corresponding feature edge (e.g., $f_1$ and $g_1$ in Figure~\ref{fig:cnn-train}). As the gradients and the features are stored in different memory locations, the gradients can use the VN for the corresponding features.

For integrity protection, MACs still need to be stored in memory. 
We customize the size of a memory block that each MAC protects
to match the data movement granularity of the accelerator.
For example, the DNN accelerator that we use for a prototype writes a 512-B chunk to memory at a time.

\vspace{-0.05in}
\subsection{\name~Instructions}

The {\name} instruction set is designed to be an extension to a DNN accelerator without changing the base instructions.
A user can choose if integrity protection is needed when initiating a session.

\noindent\textbf{\texttt{GetPK}}:
Returns the public key (PK) and the certificate (Cert). 

\noindent\textbf{\texttt{InitSession}}: 
Given a public key from a remote user, the accelerator runs a key exchange protocol to agree on
a symmetric session key and establish a secure communication channel with the user. 
The accelerator also clears all states (keys, data, and hashes), sets a new memory encryption 
key (K$_\text{MEnc}$), resets all counters to zero, and enables memory protection.
If integrity protection is enabled, memory integrity verification and hashing of instructions and their operands are also enabled.

\noindent\textbf{\texttt{SetWeight}} and \textbf{\texttt{SetInput}}:
On \textbf{\texttt{SetWeight}}, the accelerator imports encrypted weights;
these weights are decrypted with the session key (K$_\text{Session}$) and protected by
the accelerator's memory protection in DRAM.
Then, the weight counter (CTR$_\text{W}$) is incremented (see Section~\ref{sec:low_overhead_mp}).
Similarly, on \textbf{\texttt{SetInput}},
the accelerator imports the encrypted input into its protected memory, and
increments the input counter (CTR$_\text{IN}$) .
For integrity protection, the accelerator also computes the hash of the input/weights
for remote attestation.

\noindent\textbf{\texttt{ExportOutput}}:
The accelerator reads an encrypted output from DRAM,
and re-encrypts the output with K$_\text{Session}$ for the user.

\noindent\textbf{\texttt{SignOutput}}:
The accelerator computes a digital signature of the hashes of the input, output, weights, and the sequence 
of instructions/operands using its private key (SK$_\text{Accel}$).
By verifying this signature using the corresponding public key, the user can verify that the output was produced 
by the particular accelerator using the correct initial state and the correct sequence of instructions.

\noindent\textbf{\texttt{SetReadCTR}}:
To reduce overhead and allow complex compiler optimizations, \name{} relies on the host CPU to determine the VN for reading features.
This number is determined based on the network structure and scheduling and does not need to be trusted for confidentiality, as it only affects decryption.
Specifically, host CPU sets the CTR${_\text{F,R}}$ value for an address range.

In addition to the instructions listed above, the DNN may require additional preprocessing of the input data.
Those preprocessing steps can be handled by the user before sending to the accelerator.
Alternatively, \name can also handle most standard image data preprocessing, such as scaling, cropping, clipping, and reflection, by performing the data preprocessing steps as matrix multiplication.
Nvidia DALI also proposes to address the problem of the CPU bottleneck by offloading data preprocessing to the GPU.
\vspace{-0.1in}
\section{Evaluation}

\subsection{Methodology}

\noindent\textbf{FPGA Prototype --}
We implemented a prototype of \name with confidentiality-only protection (GuardNN$_\text{C}$) by adding the VN generator, encryption engines (AES-128), and a microcontroller to the CHaiDNN accelerator \cite{chaidnn}. %
We use four different DSP configurations (128, 256, 512 and 1024) with two different precisions (6-bit and 8-bit fixed point) for weights and features.
The AES engines are pipelined with a 12-cycle latency.
Because AMD Xilinx FPGAs do not currently support secure remote attestation of the bitstream, this prototype is primarily used as a functional demonstration.

\noindent\textbf{Cycle-level Simulation --}
We use cycle-level simulations to (1) compare the overhead of multiple memory protection schemes,
(2) study the overhead for a larger class of DNN models, and (3) evaluate the overhead for DNN training.
DNN accelerators are simulated using SCALE-Sim~\cite{scale-sim2}, an open-source DNN accelerator simulator from ARM research.
The memory accesses are simulated using Ramulator~\cite{ramulator} for 16GB DDR4.
\name is modeled based on Google TPU-v1~\cite{google2017tpu}, where it contains 64k processing elements (i.e., MAC units) and 24MB on-chip memory.

\noindent\textbf{Benchmarks --}
We evaluate \name~on a variety of DNN architectures --- AlexNet, VGG, GoogleNet, ResNet, MobileNet, Vision Transformer (ViT) for image classification, BERT for pretraining language models, DLRM for personalized recommendation, and wav2vec2 for learning speech representation.

\begin{table}[t]
\centering
\begin{minipage}[t]{1\linewidth}
\centering
\caption{Throughput and overhead of GuardNN FPGA prototypes --- \small{\normalfont{Throughput is reported in frames per second (fps) and overhead (\%) is calculated over CHaiDNN baseline.}}}
\label{tbl:latencyfpga}
\begin{adjustbox}{width=\linewidth}
\large
\addtolength{\tabcolsep}{-2pt} 
\begin{tabular}{@{}cccccc@{}}
\toprule
\multirow{2}{*}{\begin{tabular}[c]{@{}l@{}}Throughput \\ (Overhead) \end{tabular}} &
  \multirow{2}{*}{\# of DSPs} &
  \multicolumn{4}{c}{Network Architecture} \\ \cmidrule(l){3-6} 
          &      & AlexNet & GoogleNet            & ResNet               & VGG                  \\ \midrule
\multirow{4}{*}{\multirow{2}{*}{\begin{tabular}[c]{@{}c@{}}GuardNN$_\text{C}$ \\ (\textbf{8-bit}) \end{tabular}}} 
  & 128 & 51.5 (+0.6) & 22.1 (+0.4) & 8.1 (+1.2) & 2.5 (+0.8) \\
  & 256 & 94.5 (+0.5) & 39.4 (+0.5) & 14.6 (+1.6) & 4.8 (+0.9) \\
  & 512 & 163.6 (+0.3) & 64.7 (+1.5) & 23.7 (+1.9) & 9.0 (+0.6) \\
  & 1024 & 249.4 (\textcolor{forestgreen}{\textbf{+0.2}}) & 93.7 (+0.7) & 35.3 (\textcolor{brickred}{\textbf{+2.4}}) & 15.9 (+0.6) \\
 \midrule
 \multirow{4}{*}{\multirow{2}{*}{\begin{tabular}[c]{@{}c@{}}GuardNN$_\text{C}$ \\ (\textbf{6-bit}) \end{tabular}}} 
  & 128 & 95.2 (+0.6) & 40.4 (+0.5) & 14.9 (+1.6) & 4.8 (+0.9) \\
  & 256 & 166.3 (+0.5) & 67.2 (+0.6) & 24.6 (+2.2) & 9.1 (+0.9) \\
  & 512 & 258.1 (\textcolor{forestgreen}{\textbf{+0.3}}) & 100.2 (+0.8) & 37.6 (+2.7) & 16.5 (+0.7) \\
  & 1024 & 349.7 (\textcolor{forestgreen}{\textbf{+0.3}}) & 128.8 (+1.0) & 48.5 (\textcolor{brickred}{\textbf{+3.1}}) & 27.6 (+0.6)
          \\ \bottomrule
\end{tabular}
\end{adjustbox}
\end{minipage}
\vspace{-0.2cm}
\end{table}

\vspace{-0.05in}
\subsection{FPGA Prototype Results}

\noindent\textbf{Performance --}
Table~\ref{tbl:latencyfpga} shows the throughput and overhead for various DNN models across several different configurations on an FPGA board.
The performance overhead of {\name} is less than \textbf{3.1\%} for all eight different configurations on ImageNet.
It is worth noting that the overhead comes mainly from the limited throughput of the AES engines.
The maximum overhead among the four networks can be further reduced to \textbf{1.9\%} by increasing the number of AES engines from three (as in Table~\ref{tbl:latencyfpga}) to four.

We also studied the latency of {\name} instructions using VGG 
as an example. 
{\name} needs to perform a key exchange and load weights once per session.   
On the MicroBlaze, the \textbf{\texttt{GetPK}} and \textbf{\texttt{InitSession}} 
(specifically, the ECDHE–ECDSA key-exchange) take 23.1 ms. 
The key-exchange latency is independent of a network.
Importing (decrypting and re-encrypting) weights on \textbf{\texttt{SetWeight}} takes 19.5ms, 2.2ms, 8.0ms and 43.3ms for AlexNet, GoogleNet, ResNet, and VGG, respectively.
For each input, {\name} adds overhead to import an input and export/sign an output.
\textbf{\texttt{SetInput}} for a single input image only takes 0.1 ms.
For the 1000-class output, the \textbf{\texttt{ExportOutput}} and \textbf{\texttt{SignOutput}} take 0.01 ms and 4.8 ms, respectively.
Thus, the \name\ instructions incur negligible overhead.

\noindent\textbf{Resource Overhead --}
In our FPGA prototype with 512 DSPs and 8-bit weights/features, we use an open-source AES-128 IP core that uses 9.0K LUTs and 3.0K FFs. 
The area overhead of one AES core is 8.2\% and 2.6\% in LUTs and FFs, respectively.
Because the FPGA clock (200 MHz) is much slower than the memory bus clock, three AES engines are needed to match the memory bandwidth used by CHaiDNN.
We implemented the microcontroller as a Xilinx MicroBlaze, for which the program can fit within 256KB local memory. The microcontroller's resource usage (overhead) is 2.7K LUTs (2.5\%), 2.2K FFs (1.9\%), 64 BRAMs (11.0\%) \& 6 DSPs (0.9\%).

\vspace{-0.1in}
\subsection{ASIC Simulation Results}
\label{sec:cycle_level_sim}

\noindent\textbf{Performance --}
We study the accelerator performance for four protection schemes: no protection (NP), 
today's baseline memory protection (BP), and {\name} with confidentiality-only (GuardNN$_\text{C}$) and both confidentiality and integrity protection (GuardNN$_\text{CI}$).
GuardNN$_\text{C}$ only performs memory encryption while GuardNN$_\text{CI}$ perfoms both encryption and integrity verification.
For the baseline memory encryption, we implement the recent memory encryption engine (MEE) design from Intel~\cite{cpu_MEE} as the state-of-the-art.

\begin{figure}[t]
\subfloat[Inference.]{
\begin{minipage}{\linewidth}
    \centering
        \includegraphics[width=3.3in]{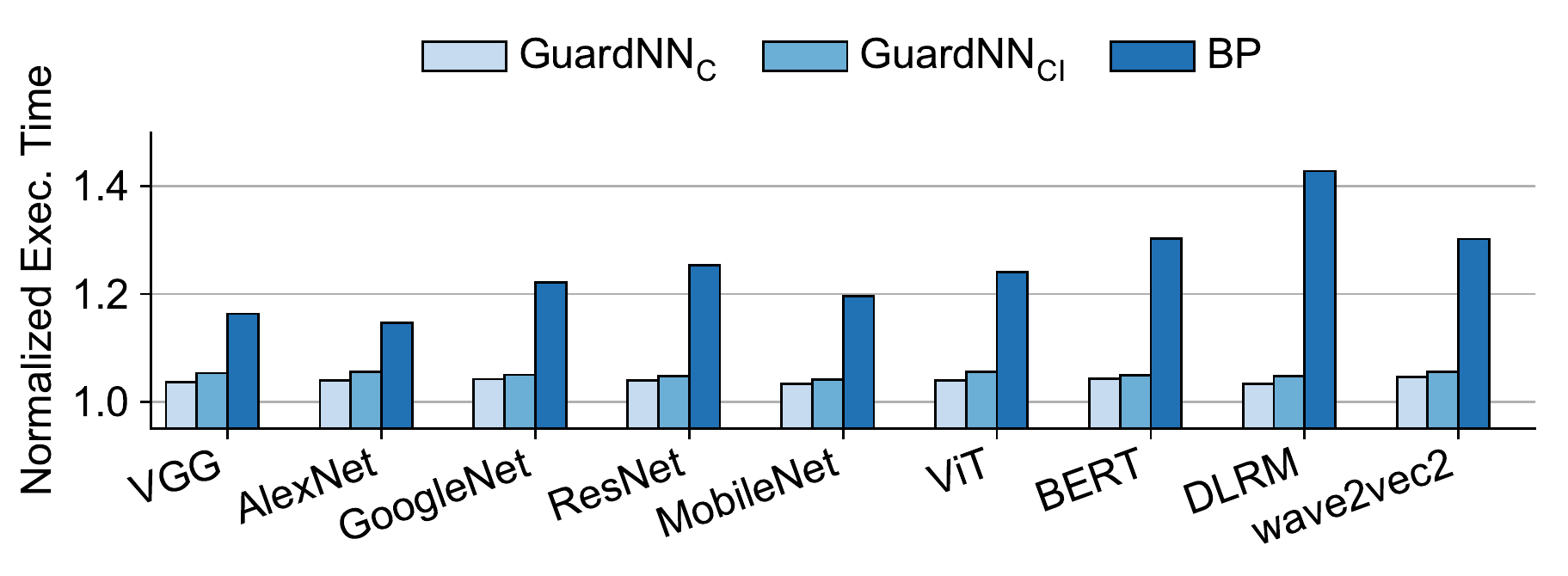}
        \vspace{-0.2cm}
        \label{fig:lat_inf}
\end{minipage}}
\vspace{-0.4cm}
\subfloat[Training.]{
\begin{minipage}{\linewidth}
    \centering
        \includegraphics[width=3.3in]{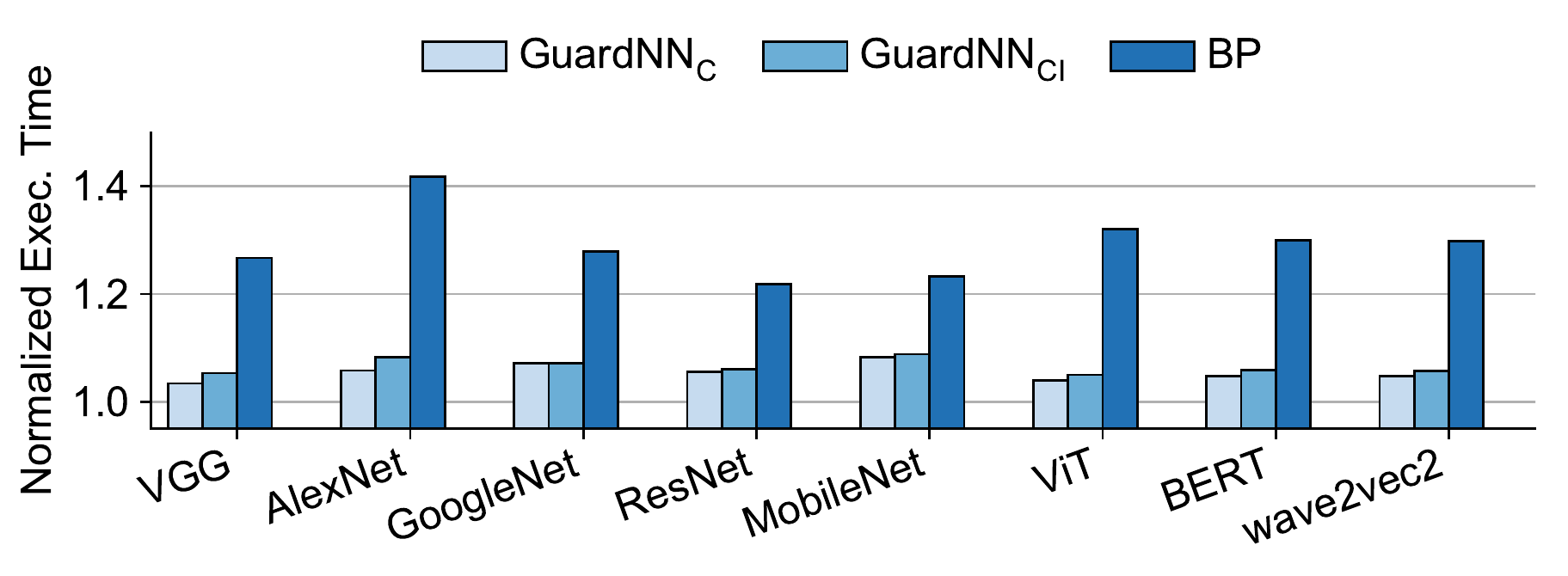}
        \vspace{-0.2cm}
        \label{fig:lat_train}
\end{minipage}}
\caption{The normalized execution time of the DNN inference and training on different networks models.}
\label{fig:lat}
\end{figure}

As the throughput of a DNN accelerator is often limited by the memory bandwidth,
we first compare the memory traffic increase of BP and GuardNN.
The memory traffic increase is defined as the ratio between the total number of memory accesses with and without memory protection.
BP increases memory accesses by 35.3\% on average for inference and by 37.8\% for training.
The memory traffic increase is larger for training because the training process accesses more data and has more frequent cache evictions in the \VN/MAC cache.
GuardNN$_\text{CI}$ increases the memory traffic by \textbf{2.4\%} and \textbf{2.3\%} on average for inference and training, respectively.

Figure~\ref{fig:lat} shows the performance of the baseline protection and the DNN-specific memory protection.
BP is 1.25$\times$ and 1.29$\times$ slower than no protection on average for inference and training.
For inference and training, both GuardNN$_\text{C}$ and GuardNN$_\text{CI}$ show much lower performance overhead than \texttt{BP}. The average overhead of GuardNN$_\text{CI}$ is \textbf{1.05\%} for inference and \textbf{1.07\%} for training. 
GuardNN$_\text{C}$ further reduces the overhead to \textbf{1.04\%} for inference and \textbf{1.05\%} for training.
The results demonstrate that {\name} can support secure DNN processing with negligible overhead over the baseline accelerator with no protection.

\noindent\textbf{ASIC Power/Area Overhead --} %
The power and area overhead of {\name} is expected to be low for an ASIC design. 
The additional area mainly comes from
the AES engines, which are used for encryption and integrity verification. 
A previous study~\cite{asic_aes_vlsi19} shows that a low-power AES engine only consumes $0.0031~\text{mm}^2$ in area and 3.85 mW in power, while achieving a 991 Mbps throughput at 875 MHz in 28nm ASIC. 
In contrast, the area and power consumption of TPU-v1~\cite{google2017tpu} (also in 28nm) are $331~\text{mm}^2$ and 75 W, respectively. Notably, TPU-v1 runs at 700 MHz and has a peak memory bandwidth of 272 Gbps.
In order to match the memory bandwidth of TPU-v1, we can instantiate 344 AES engines, which only result in 0.3\% area and 1.8\% power overhead. We can also use a smaller number of high-performance AES engines, which will likely have similar overall overhead.

\begin{table*}[t]
\large
\centering
\caption{Comparison between different privacy-preserving ML approaches --- \small{\normalfont{The throughput is measured in giga operations per second (GOPs) and the energy efficiency is reported in giga operations per second per Watt (GOPs/W). For GuardNN accelerators, we show the number of PEs, the size of on-chip SRAM, and the clock frequency. The power of the \name is estimated based on TPU-v1.}}}
\label{tab:comparision}
\begin{adjustbox}{width=0.9\linewidth}
\begin{tabular}{@{}ccccccc@{}}
\toprule
\multicolumn{2}{c}{\multirow{2}{*}{Metrics}}    & \multicolumn{5}{c}{Methods}       \\ \cmidrule(l){3-7} 
\multicolumn{2}{c}{} &
  \begin{tabular}[c]{@{}c@{}}CPU TEE\\ (Simulated)\end{tabular} &
  \begin{tabular}[c]{@{}c@{}}DELPHI\\ MPC~\cite{delphi}\end{tabular} &
  \begin{tabular}[c]{@{}c@{}}CrypTFLOW2\\ MPC~\cite{cryptflow2}\end{tabular} &
  \begin{tabular}[c]{@{}c@{}}GuardNN$_\text{CI}$\\ (Simulated)\end{tabular} &
  \begin{tabular}[c]{@{}c@{}}GuardNN$_\text{C}$\\ (FPGA)\end{tabular} \\ \midrule
\multicolumn{2}{c}{Hardware} &
  \begin{tabular}[c]{@{}c@{}} CPU\\ 1 core@3.0 GHz\end{tabular} &
  \begin{tabular}[c]{@{}c@{}}Intel Xeon \\ 4 cores@3.7 GHz\end{tabular} &
  \begin{tabular}[c]{@{}c@{}}Intel Xeon\\ 4 cores@3.7 GHz\end{tabular} &
  \begin{tabular}[c]{@{}c@{}}64k PEs/24 MBs\\@0.7 GHz\end{tabular} &
  \begin{tabular}[c]{@{}c@{}}512 PEs/3 MBs \\@0.2 GHz \end{tabular} \\ \midrule
\multirow{2}{*}{Workloads}   & Network     & VGG-16 & ResNet-32 & ResNet-32 & VGG-16 & VGG-16 \\ \cmidrule(lr){2-2}
                         & Dataset             & ImageNet & CIFAR-100 & CIFAR-100  & ImageNet & ImageNet \\ \midrule
\multirow{2}{*}{Perf.}   & Throughput   & 0.81 & 0.02 & 0.18 & 3221.57 & 139.23 \\ \cmidrule(lr){2-2}
                         & Overhead ($\times$)            & 1.61 &  $\sim$1000    & $\sim$100 & 1.05 & 1.01   \\ \midrule
\multirow{2}{*}{Energy}  & Power (W)            & $\sim$60 &  130 & 130 & $\sim$40  & $\sim$15 \\ \cmidrule(lr){2-2}
                         & Efficiency   & 0.01 & 0.002 & 0.0001 & 80.5  & 9.3  \\ \midrule
\multirow{2}{*}{TCB}     & Components       & CPU & MPC protocol       & MPC protocol & Accelerator  & Accelerator \\ \cmidrule(lr){2-2}
                         & Lines of code         & Millions~\cite{keystone} & 35.1k          & 53.7k          & 10-100s of thousands & 21.8k \\ \bottomrule
\end{tabular}
\end{adjustbox}
\end{table*}

\subsection{Comparison with Alternatives}

Table~\ref{tab:comparision} compares {\name} with other approaches for privacy-preserving deep learning.
For a CPU TEE, the table shows the performance for a simulated CPU TEE with unlimited protected memory
to represent the ideal case of CPU TEEs.
Today's Intel SGX limits its protected memory to 128 MBs, which leads to 5-20$\times$ performance overhead for DNNs \cite{Vessels2020}. 

GuardNN$_\text{CI}$ provides confidentiality and integrity guarantees for VGG with only 5\% overhead, while the simulated CPU TEE adds more than 60\% overhead on the same benchmark.
By leveraging hardware-based protection and high performance accelerators, \name~achieves three orders of magnitude higher performance and energy efficiency compared to alternatives.
As discussed in Section~\ref{sec:arch}, \name~also has a small TCB thanks to the simplicity of DNN accelerator design and protection mechanisms.
The lines of code (LoC) for GuardNN prototype is 21.8k in total --- 9k LoC for the baseline accelerator, 8.3k LoC for the customized protection, and 4.5k LoC for new instructions (firmware on a microcontroller).
MPC-based approaches also have a relatively small size of TCB and can be implementation with tens of thousands of LoC. However, their performance overhead is significantly higher than {\name}.
While MPC-based approaches offer secure and easy-to-deploy options for less performance-demanding use cases, secure accelerators appear to be the most promising solution for large-scale, high-throughput use cases.
\vspace{-0.05in}
\section{Related Work}
\textbf{Privacy-Preserving Deep Learning --} \name~provides hardware-based protection for DNN inference and training in an untrusted environment.
Alternatively, homomorphic encryption (HE) and secure multi-party computation (MPC) can provide stronger protection on today's hardware by performing
all computations in an encrypted format. 
While HE and MPC provide strong cryptographic guarantees without trusting remote hardware or software, they come with
significant overhead compared to the baseline with no protection \cite{CryptoNets_FHME, delphi, cryptflow2020, wagh2021falcon, AriaNN, HE_large_model}.
A recent work~\cite{cheetah} proposes to reduce the latency of HE-based DNN inference to hundreds of milliseconds using specialized hardware.
\name~provides a design point that provides hardware-based security with much 
higher performance compared to the HE/MPC-based solutions.

TEEs provide hardware-protected execution environments where
confidentiality and integrity are ensured even under an untrusted OS or physical attacks.
Recent studies showed that DNN computations can be protected using Intel SGX~\cite{slalom, Vessels2020, occlumency} but with non-trivial overhead of memory protection in SGX. 
Today's processor-based TEEs are also limited by the performance of a general-purpose processor.
Recent studies~\cite{graviton, HIX} proposes to extend today's TEE by including a GPU. 
The GPU-based TEEs enable much higher DNN performance compared to general-purpose processors, but require 
both a CPU and a GPU to be protected inside a TEE.
Telekine~\cite{Telekine} further improves the security of GPU TEEs by translating the application's GPU API calls into data-oblivious commands.

Recent work \cite{htee-sp20, NPUFort_secure_hardware, FPGA_TEE} proposes to build FPGA/ASIC TEEs as accelerators are often far more energy-efficient than GPUs and widely used for high-throughput tasks such as inference.
TNPU~\cite{TNPU} concurrently proposes a tree-less off-chip memory protection for DNN accelerators, similar to our DNN-specific memory protection.
{\name} allows a smaller TCB and lower overhead by carefully desining its instruction set for an untrusted host and 
an option for confidentiality-only protection.

\name~proposes a new approach to enable secure DNN computation using accelerators and shows that secure accelerators have a potential to provide higher security with a negligible performance and area overhead compared to the general-purpose platforms
by customizing its architecture and protection for DNNs.

\textbf{Side-channel Attacks and Protection --}
A variety of side-channel attacks have been shown to work against DNN accelerators. Memory and timing side-channels have been used to infer the network structure of an accelerator with encrypted weights \cite{recnn, recnn_journal}. 
\name~has a fixed memory access pattern and execution time, and is 
secure against memory and timing side-channels.
Physical side-channel attacks such as power and electromagnetic side-channel attacks have been used to retrieve the input image~\cite{powersidechan} or recover the network topology and weights~\cite{usenixem}.
If strong protection against power and EM side-channel attacks is necessary, \name~needs to be extended with additional countermeasures.
\vspace{-0.08in}
\section{Conclusion}

This paper proposes a secure DNN accelerator, named
\name. %
Application-specific accelerators provide strong isolation from
a CPU with complex software stack and also enable protection to be
customized for DNNs to improve both security and performance.
An FPGA prototype shows that the \name can protect common DNN models
with minimal ($\sim$3\%) performance overhead. 
\vspace{-0.08in}
\section{Acknowledgment}
We thank the anonymous reviewers for their constructive feedback.
At Cornell, Weizhe Hua and Muhammad Umar are supported in part by NSF Award CCF-2007832, ECCS-1932501, and CCF-2118709. Weizhe Hua is also supported in part by the Facebook fellowship.

\bibliographystyle{ieeetr}
{\bibliography{refs}}

\end{document}